\documentclass[conference,letterpaper]{IEEEtran}
\IEEEoverridecommandlockouts

\usepackage{amsmath,amssymb,amsfonts}
\usepackage{algorithmic}
\usepackage{graphicx}
\usepackage{textcomp}
\usepackage{xcolor}
\usepackage{cite}

\usepackage[a4paper,
            left=0.625in,
            right=0.625in,
            top=0.75in,
            bottom=1in]{geometry}

\usepackage{amsmath,amssymb,amsfonts}
\usepackage{algorithmic}
\usepackage{graphicx}
\usepackage{textcomp}
\usepackage{xcolor}
\usepackage{gensymb}

\usepackage{enumitem}
\usepackage{graphicx}
\usepackage{setspace}
\usepackage{titlesec} 
\usepackage{xcolor}
\usepackage{amsmath}
\usepackage{array}
\usepackage{booktabs}  
\usepackage{amssymb}
\usepackage{float}
\usepackage[normalem]{ulem}
\usepackage{cite}

\usepackage{tabularx}
\usepackage{amssymb}
\usepackage{tikz}
\usetikzlibrary{arrows.meta, positioning}
\usepackage{amsmath}
\usepackage{algorithm}

\usepackage{multicol}
\usepackage{blindtext}
\usepackage{longtable}
\usepackage{lipsum} 
\expandafter\def\expandafter\normalsize\expandafter{%
    \normalsize%
    \setlength\abovedisplayskip{0pt}%
    \setlength\belowdisplayskip{8pt}%
    \setlength\abovedisplayshortskip{-8pt}%
    \setlength\belowdisplayshortskip{2pt}%
}
\usepackage[font=small,skip=0pt]{caption}
\usepackage{lipsum}
\usepackage{comment}

\usepackage[normalem]{ulem}

\usepackage{algorithmic}

\usepackage{mathtools,amssymb,lipsum}

\usepackage{cuted}
\usepackage{stfloats}

\usepackage[a4paper,
            left=0.625in,
            right=0.625in,
            top=0.75in,
            bottom=1in]{geometry}
            
\usepackage[normalem]{ulem}

\usepackage{bm}       

 \usepackage{booktabs}
\usepackage{makecell}
 \usepackage{amssymb} 

\begin{document}

\title{SAC-Controlled RIS-Assisted Monopulse Radar for Multipath-Robust Vital-Sign Localization}

\author{Jafar~Norolahi,~Alireza~Vahid
\thanks{J. Norolahi and A. Vahid are with the Department of Electrical and Microelectronic Engineering at Rochester Institute of Technology, Rochester, NY 14623, USA. Emails: {jn3211@rit.edu, arveme@rit.edu}.
}
}

\maketitle

\begin{abstract}
Contactless radar sensing offers a promising approach to continuous vital-sign
monitoring. However, indoor multipath can distort direction estimates and
generate ghost responses. This paper presents a reconfigurable intelligent
surface (RIS)-assisted monopulse radar whose RIS phase configuration is
controlled by a soft actor-critic agent to improve angular localization.
The proposed framework combines simultaneous sum and difference beams with a
programmable reflected path, enabling a single radar node to adapt the
propagation environment without relying on multiple synchronized sensors. A
compact multipath model formulates RIS control as a continuous-action
reinforcement-learning problem whose reward penalizes bearing errors and
misleading multipath reinforcement. Numerical results compare the proposed
method with conventional monopulse processing without an RIS, a
frequency-modulated continuous-wave range-angle fast Fourier transform
baseline, random RIS phase configurations, and deterministic RIS controllers.
The evaluation considers different target separations, signal-to-noise ratio
levels, and multipath conditions.
\end{abstract}

\section{Introduction}
\label{sec:introduction}

Continuous monitoring of respiration and other vital signs is important in hospitals, assisted-living facilities, and smart-home healthcare systems. Conventional monitoring devices, including adhesive electrodes, chest belts, and wearable photoplethysmography sensors, can provide reliable measurements but require direct physical contact and regular user cooperation. Their long-term use may be limited by skin irritation, battery charging, discomfort, and inconsistent adherence, particularly for elderly, pediatric, or bedridden patients \cite{Zhang2023Photonic,Shi2020Dataset,10299799}. Camera-based monitoring provides a contactless alternative, but its performance is sensitive to illumination, occlusion, and patient orientation and may also raise privacy concerns \cite{Villarroel2019Noncontact}. These limitations have motivated the development of radio-frequency sensing systems that can monitor physiological motion without requiring a wearable device or a clear optical view of the subject.

Radar-based vital-sign sensing detects the small periodic displacement of the chest produced by respiration and heartbeat. Continuous-wave and frequency-modulated continuous-wave radars have consequently been investigated for contactless physiological monitoring, patient detection, and indoor tracking \cite{Zhang2021MTrack,10058975,10216298}. However, reliable operation becomes difficult when multiple subjects or strong reflectors occupy the same environment. Reflections from walls, furniture, and medical equipment may create ghost targets or distort the estimated direction of arrival. Moreover, conventional radar deployments may contain blind regions caused by blockage, unfavorable target orientation, or weak propagation paths. Multiple spatially separated radar nodes can improve localization coverage, but they increase hardware cost and require synchronization and data fusion among the sensing nodes.

Monopulse processing provides an attractive alternative for reducing this infrastructure requirement. A monopulse radar forms a sum beam and one or more difference beams simultaneously. The ratios between the received difference- and sum-channel signals provide angular error information from the same observation interval, allowing the target bearing to be estimated without mechanically scanning the entire surveillance region \cite{10552089,10857303,Norolahi2022Beamforming}. This property is particularly useful for vital-sign monitoring because the target response may vary over time due to posture changes and small body movements. Nevertheless, the accuracy of the monopulse ratio remains dependent on the propagation environment. Multipath components entering the sum and difference channels with different phases can bias the angular estimate, especially when the direct target reflection is weak.

\begin{figure}[t]
    \centering
    \includegraphics[
        width=0.9\columnwidth
    ]{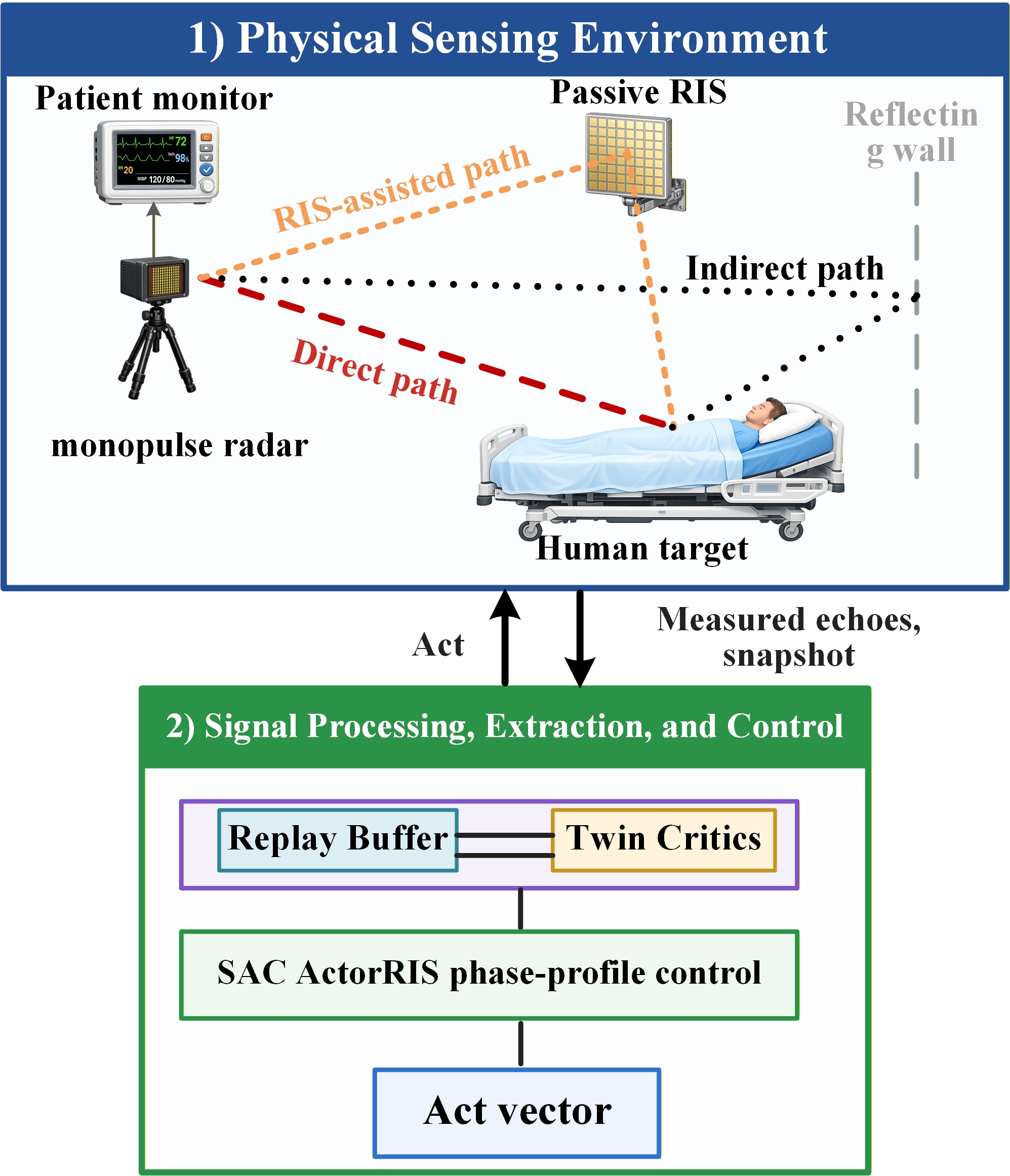}
    \caption{Closed-loop RIS-assisted monopulse localization.
    The SAC agent is trained offline using simulated rewards.
    During deployment, the frozen actor maps monopulse
    observations to RIS phase shifts without access to
    ground-truth target angles.}
    \label{fig:conference_sac_framework}
\end{figure}

A reconfigurable intelligent surface (RIS) can provide an additional controllable propagation path between the radar and the monitored subject. An RIS consists of programmable reflecting elements whose phase responses can be adjusted to redirect or focus incident electromagnetic energy \cite{9994338,10216298,Norolahi2025STARRIS,Norolahi2026DistributedRIS}. Proper RIS configuration can strengthen the desired target response, improve illumination of partially blocked regions, and reduce the effect of unfavorable multipath propagation \cite{10530308,10092469}. However, selecting the RIS phases is challenging because the phase-control space grows rapidly with the number of reflecting elements, while the best configuration changes with the target position and surrounding reflectors. Conventional beam steering based only on a nominal target direction may therefore be insufficient in a dynamic multipath environment.

Deep reinforcement learning offers a model-free mechanism for adapting high-dimensional RIS controls according to radar observations \cite{Norolahi2023MLB5G}. In particular, the soft actor-critic (SAC) algorithm supports continuous actions and employs entropy regularization to balance exploration and exploitation during training \cite{9718218,10345616,9923640}. These characteristics make SAC suitable for optimizing continuous RIS phase shifts when an analytical relationship between the phase configuration and the resulting monopulse error is difficult to obtain. Once trained, the policy can directly determine the RIS configuration from the observed monopulse measurements without solving a computationally intensive optimization problem for every sensing interval\cite{Hoang2026AdaptiveBeampatterns}.

As shown in Fig. \ref{fig:conference_sac_framework}, this paper investigates SAC-controlled RIS-assisted monopulse radar for angular localization of vital-sign targets in multipath indoor environments. The scope is intentionally centered on bearing estimation and RIS phase adaptation rather than the complete physiological-signal reconstruction process. The main contributions are summarized as follows:

\begin{itemize}
    \item We develop a compact RIS-assisted monopulse signal model that incorporates the direct target reflection, an RIS-controlled reflection, and multipath-induced ghost components in the sum and difference channels.

    \item We formulate RIS phase adaptation as a continuous-control problem and employ SAC to determine the reflecting-element phases from monopulse observations. The reward is designed to reduce angular localization error while preventing the RIS from reinforcing misleading multipath responses.

    \item We characterize the bearing-offset estimation performance
    of the proposed controller for two closely spaced targets over
    true azimuth separations from \(0.1^\circ\) to \(1^\circ\), and
    quantify the corresponding absolute angular error.
\end{itemize}

\section{System Model}
\label{sec:system_model}

We consider a monostatic monopulse radar equipped with an
\(M\)-element planar array and assisted by one passive
reconfigurable intelligent surface (RIS) containing \(N\)
programmable reflecting elements. Here,
\(M=M_rM_c\) and \(N=N_rN_c\), where \(M_r,M_c\) and \(N_r,N_c\)
denote the numbers of array rows and columns, respectively.
The radar monitors \(K\) subjects in an indoor environment containing
a direct propagation path, an RIS-assisted path, and
\(G\) wall-induced ghost paths. In the simulations,
\(M=N=40\times40=1600\) and \(K=2\).
The conference study focuses on angular localization; range and
physiological-frequency estimation are reserved for the extended
journal version.

Let the radar array be located at the origin, and let the direction
of the \(k\)-th target be represented by its azimuth
\(\theta_k\) and elevation \(\phi_k\). The corresponding unit
direction vector is

\begin{align}
\mathbf{u}(\theta_k,\phi_k)
=
\begin{bmatrix}
\cos\phi_k\cos\theta_k \\
\cos\phi_k\sin\theta_k \\
\sin\phi_k
\end{bmatrix}.
\label{eq:conference_direction}
\end{align}

If \(\mathbf{r}_m\in\mathbb{R}^{3\times1}\) denotes the position
of the \(m\)-th radar element, the \(m\)-th entry of the array
steering vector is

\begin{align}
\left[\mathbf{a}(\theta_k,\phi_k)\right]_m
=
\exp\!\left(
j\frac{2\pi}{\lambda}
\mathbf{r}_m^{T}\mathbf{u}(\theta_k,\phi_k)
\right),
\label{eq:conference_steering}
\end{align}

where \(j=\sqrt{-1}\), \(\lambda=c/f_c\) is the wavelength,
\(c\) is the speed of light, and \(f_c\) is the carrier
frequency. The target manifold matrix is

\begin{align}
\mathbf{A}
=
\left[
\mathbf{a}(\theta_1,\phi_1),
\ldots,
\mathbf{a}(\theta_K,\phi_K)
\right]
\in\mathbb{C}^{M\times K}.
\label{eq:conference_manifold}
\end{align}

The direct-path coefficient of target \(k\) during dwell \(t\)
is represented by

\begin{align}
d_{k,t}
=
\alpha_{k,t}e^{j\xi_{k,t}},
\label{eq:conference_direct_coefficient}
\end{align}

where \(\alpha_{k,t}\) contains the blocked-path attenuation
and target reflectivity, and \(\xi_{k,t}\) is the corresponding
propagation phase. The coefficient is treated as constant over one
coherent radar dwell. Stacking the coefficients gives

\begin{align}
\mathbf{d}_t
=
[d_{1,t},\ldots,d_{K,t}]^{T}.
\end{align}

A wall or another large reflector can create virtual target
directions. For the \(g\)-th ghost path,
\(\mathbf{A}^{(g)}\in\mathbb{C}^{M\times K}\) denotes the
corresponding ghost manifold and
\(\mathbf{g}^{(g)}_t\in\mathbb{C}^{K\times1}\) contains its
complex path coefficients.

The RIS phase-control matrix is

\begin{align}
\boldsymbol{\Phi}_t
=
\operatorname{diag}
\left(
e^{j\varphi_{1,t}},
\ldots,
e^{j\varphi_{N,t}}
\right),
\qquad
\varphi_{n,t}\in[-\pi,\pi].
\label{eq:conference_ris_matrix}
\end{align}

Let \(\mathbf{H}_{r}\in\mathbb{C}^{M\times N}\) denote the
effective RIS-to-radar return channel, and let
\(\mathbf{B}\in\mathbb{C}^{N\times K}\) represent the
target-to-RIS spatial response. The RIS-assisted return is

\begin{align}
\mathbf{x}^{\mathrm{RIS}}_t
=
\mathbf{H}_{r}
\boldsymbol{\Phi}_t
\mathbf{B}
\mathbf{q}_t,
\label{eq:conference_ris_return}
\end{align}

where \(\mathbf{q}_t\in\mathbb{C}^{K\times1}\) contains the
effective two-way RIS-assisted path coefficients.

The complete post-dwell array snapshot is

\begin{align}
\mathbf{x}_t
={}&
\underbrace{\mathbf{A}\mathbf{d}_t}_{\text{direct paths}}
+
\underbrace{
\sum_{g=1}^{G}
\mathbf{A}^{(g)}
\mathbf{g}^{(g)}_t
}_{\text{ghost paths}}
\nonumber\\
&+
\underbrace{
\mathbf{H}_{r}
\boldsymbol{\Phi}_t
\mathbf{B}
\mathbf{q}_t
}_{\text{RIS-assisted paths}}
+
\mathbf{n}_t,
\label{eq:conference_snapshot}
\end{align}

where

\begin{align}
\mathbf{n}_t
\sim
\mathcal{N}_{\mathbb{C}}
\left(
\mathbf{0},
\sigma_n^2\mathbf{I}_M
\right),
\end{align}

with \(\sigma_n^2\) denoting the post-dwell noise variance
and \(\mathbf{I}_M\) the \(M\times M\) identity matrix.

\subsection{Monopulse Bearing Estimation}
\label{subsec:conference_monopulse}

At dwell \(t\), the radar maintains a prior azimuth direction
\(\bar{\theta}_{k,t}\) for target \(k\).
The elevation is fixed at the known value \(\phi_0\) in the
present simulations. The normalized sum-channel weight is
\begin{align}
\mathbf{w}_{\Sigma,k,t}
=
\frac{
\mathbf{a}(\bar{\theta}_{k,t},\phi_0)
}{
\left\|
\mathbf{a}(\bar{\theta}_{k,t},\phi_0)
\right\|_2
}.
\label{eq:conference_sum_weight}
\end{align}

Let

\begin{align}
\mathbf{a}'_{\theta,k,t}
=
\left.
\frac{\partial\mathbf{a}(\theta,\phi_0)}
{\partial\theta}
\right|_{\theta=\bar{\theta}_{k,t}} .
\end{align}

The component of this derivative parallel to the sum steering
vector is removed according to

\begin{align}
\mathbf{q}_{\theta,k,t}
=
\mathbf{a}'_{\theta,k,t}
-
\mathbf{a}_{k,t}
\frac{
\mathbf{a}_{k,t}^{H}\mathbf{a}'_{\theta,k,t}
}{
\mathbf{a}_{k,t}^{H}\mathbf{a}_{k,t}
},
\label{eq:conference_diff_theta_raw}
\end{align}

where
\(\mathbf{a}_{k,t}=\mathbf{a}(\bar{\theta}_{k,t},\phi_0)\).
The normalized azimuth-difference weight is

\begin{align}
\mathbf{w}_{\Delta,k,t}
=
\frac{
\mathbf{q}_{\theta,k,t}
}{
\left\|
\mathbf{q}_{\theta,k,t}
\right\|_2
}.
\label{eq:conference_difference_normalization}
\end{align}

The sum and difference outputs are

\begin{align}
x_{\Sigma,k,t}
&=
\mathbf{w}_{\Sigma,k,t}^{H}\mathbf{x}_t,
\label{eq:conference_sum_output}
\\
x_{\Delta,k,t}
&=
\mathbf{w}_{\Delta,k,t}^{H}\mathbf{x}_t.
\label{eq:conference_theta_output}
\end{align}

The locally calibrated monopulse slope is

\begin{align}
\kappa_{\theta,k,t}
=
\operatorname{Re}
\left\{
\frac{
\mathbf{w}_{\Delta,k,t}^{H}
\mathbf{a}'_{\theta,k,t}
}{
\mathbf{w}_{\Sigma,k,t}^{H}
\mathbf{a}_{k,t}
}
\right\}.
\label{eq:conference_monopulse_slope}
\end{align}

For a sufficiently small angular offset, the azimuth correction is

\begin{align}
\delta\theta_{k,t}
=
\operatorname{clip}
\left(
\frac{
\operatorname{Re}
\left\{
x_{\Delta,k,t}/
(x_{\Sigma,k,t}+\epsilon)
\right\}
}{
\kappa_{\theta,k,t}
},
-2^{\circ},
2^{\circ}
\right).
\label{eq:conference_theta_ratio}
\end{align}
where \(\epsilon>0\) prevents division by a small sum-channel
output and \(\operatorname{clip}(x,l,u)\) limits \(x\) to
\([l,u]\). The resulting azimuth estimate is

\begin{align}
\widehat{\theta}_{k,t}
=
\bar{\theta}_{k,t}
+
\delta\theta_{k,t}.
\label{eq:conference_theta_estimate}
\end{align}

Only azimuth is estimated in the present study; elevation
estimation is outside its scope. Ghost components generally arrive
from directions different from the target direction and perturb
both \(x_{\Sigma,k,t}\) and \(x_{\Delta,k,t}\). The RIS phases are
therefore selected to strengthen the desired target contribution
while reducing ghost-induced monopulse bias.

\section{SAC-Based RIS Phase Control}
\label{sec:sac_design}

The implemented controller is a one-step entropy-regularized SAC
problem: one calibration state produces one RIS configuration,
which is evaluated over one radar dwell.

At dwell \(t\), where \(t\) denotes the dwell index, the
calibration subsystem provides noisy estimates of the
target-enhancement phase \(\widehat{\eta}_{T,t}\), ghost-control
phase \(\widehat{\eta}_{G,t}\), ghost-control magnitude
\(\widehat{\beta}_{G,t}\), and ghost angular offset
\(\widehat{\delta}_{G,t}\). The first two quantities specify
the phase offsets required for target enhancement and ghost
suppression, respectively; \(\widehat{\beta}_{G,t}\in[0,1]\)
specifies the estimated ghost-control strength, and
\(\widehat{\delta}_{G,t}\) is the estimated angular displacement
of the ghost from the target. For the two tracked targets, the
state is

\begin{align}
\mathbf{s}_t
=
\Big[
&
\cos\widehat{\eta}_{T,t},
\sin\widehat{\eta}_{T,t},
\cos\widehat{\eta}_{G,t},
\sin\widehat{\eta}_{G,t},
\widehat{\beta}_{G,t},
\nonumber\\
&
\rho^{\mathrm{R}}_{1,t},
\rho^{\mathrm{R}}_{2,t},
\rho^{\mathrm{I}}_{1,t},
\rho^{\mathrm{I}}_{2,t},
\nonumber\\
&
\zeta_{P,t},
\zeta_{\Delta,t},
\zeta_{\mathrm{SNR},t},
\zeta_{G,t}
\Big]^{T}
\in\mathbb{R}^{13}.
\label{eq:conference_state}
\end{align}

The monopulse ratios and their bounded real and imaginary
components are

\begin{align}
\rho_{k,t}
&=
\frac{x_{\Delta,k,t}}
{x_{\Sigma,k,t}+\epsilon},
\qquad k\in\{1,2\},
\\
\rho^{\mathrm{R}}_{k,t}
&=
\operatorname{clip}
\left(\operatorname{Re}\{\rho_{k,t}\},-1.5,1.5\right),
\\
\rho^{\mathrm{I}}_{k,t}
&=
\operatorname{clip}
\left(\operatorname{Im}\{\rho_{k,t}\},-1.5,1.5\right),
\end{align}
where \(\epsilon>0\) prevents numerical division by zero and
\(\operatorname{clip}(x,l,u)\) limits \(x\) to \([l,u]\). The
remaining normalized state features are

\begin{align}
\zeta_{P,t}
&=
\frac{1}{2}
\operatorname{clip}
\left[
\log_{10}
\left(
\frac{|x_{\Sigma,2,t}|^2+\epsilon}
{|x_{\Sigma,1,t}|^2+\epsilon}
\right),
-2,2
\right],
\\
\zeta_{\Delta,t}
&=
\frac{
\left|
\bar{\theta}_{2,t}-\bar{\theta}_{1,t}
\right|
}{
1.2^{\circ}
},
\\
\zeta_{\mathrm{SNR},t}
&=
\frac{
\mathrm{SNR}_{t,\mathrm{dB}}-5
}{
10
},
\\
\zeta_{G,t}
&=
\frac{
\operatorname{rad2deg}
\left(\widehat{\delta}_{G,t}\right)
}{
6
}.
\label{eq:conference_state_features}
\end{align}

Here, \(\zeta_{P,t}\) is the normalized sum-channel power
imbalance, \(\zeta_{\Delta,t}\) is the normalized prior angular
separation, \(\zeta_{\mathrm{SNR},t}\) is the normalized
post-dwell SNR, and \(\zeta_{G,t}\) is the normalized ghost
offset. The constants \(1.2^{\circ}\), \(5~\mathrm{dB}\),
\(10~\mathrm{dB}\), and \(6^{\circ}\) are the fixed normalization
constants used during training. The state contains no true target
angles.

The SAC actor first generates

\begin{align}
\mathbf{v}_t
&=
\boldsymbol{\mu}_{\psi}(\mathbf{s}_t)
+
\boldsymbol{\sigma}_{\psi}(\mathbf{s}_t)
\odot
\boldsymbol{\varepsilon}_t,
\\
\mathbf{u}_t
&=
\tanh(\mathbf{v}_t),
\qquad
\boldsymbol{\varepsilon}_t
\sim
\mathcal{N}(\mathbf{0},\mathbf{I}_5),
\label{eq:conference_sac_action}
\end{align}

where \(\psi\) denotes the actor parameters,
\(\boldsymbol{\mu}_{\psi}\) and
\(\boldsymbol{\sigma}_{\psi}\) are the
five-dimensional Gaussian mean and standard-deviation
outputs, \(\odot\) denotes element-wise multiplication,
\(\mathbf{I}_5\) is the \(5\times5\) identity matrix, and
\(\tanh(\cdot)\) is applied element-wise. The first two
action pairs are projected onto the unit circle:

\begin{align}
[c_{T,t},s_{T,t}]
&=
\frac{[u_{1,t},u_{2,t}]}
{\|[u_{1,t},u_{2,t}]\|_2},
\\
[c_{G,t},s_{G,t}]
&=
\frac{[u_{3,t},u_{4,t}]}
{\|[u_{3,t},u_{4,t}]\|_2},
\\
b_t
&=
\operatorname{clip}(u_{5,t},-1,1).
\end{align}

If either phase pair has a near-zero norm, it is replaced by
\([1,0]\) before decoding. The five-dimensional structured action
is

\begin{align}
\mathbf{a}_t
=
[c_{T,t},s_{T,t},c_{G,t},s_{G,t},b_t]^{T}.
\label{eq:conference_action}
\end{align}

The pair \([c_{T,t},s_{T,t}]\) encodes the
target-enhancement phase, \([c_{G,t},s_{G,t}]\) encodes the
ghost-control phase, and \(b_t\) encodes the ghost-control
strength. These quantities are decoded as

\begin{align}
\eta_{T,t}
&=
\operatorname{atan2}(s_{T,t},c_{T,t}),
\\
\eta_{G,t}
&=
\operatorname{atan2}(s_{G,t},c_{G,t}),
\\
\beta_{G,t}
&=
\frac{b_t+1}{2}
\in[0,1].
\label{eq:conference_action_decoding}
\end{align}

Here, \(\operatorname{atan2}(\cdot,\cdot)\) recovers a phase
from its sine--cosine pair, and \(\beta_{G,t}\) determines the
fraction of the ghost-control columns that combine coherently.
The remaining columns use alternating \(0\) and \(\pi\) phase
offsets; therefore, every RIS element remains unit modulus.

These controls are expanded into the complete \(N\)-element RIS
phase mask. 
During offline training, the simulator provides the true target
azimuths. The target-angle RMSE and separation error are

\begin{align}
e_{\mathrm{tar},t}
&=
\sqrt{
\frac{1}{K}
\sum_{k=1}^{K}
\left[
\operatorname{rad2deg}
\left(
\widehat{\theta}_{k,t}-\theta_{k,t}
\right)
\right]^2
},
\\
e_{\mathrm{sep},t}
&=
\left|
\operatorname{rad2deg}
\left(
\left|
\widehat{\theta}_{2,t}-\widehat{\theta}_{1,t}
\right|
-
\left|
\theta_{2,t}-\theta_{1,t}
\right|
\right)
\right|.
\label{eq:conference_angular_error}
\end{align}

Here, \(\operatorname{rad2deg}(\cdot)\) converts radians to
degrees, \(e_{\mathrm{tar},t}\) measures the individual-target
bearing error, and \(e_{\mathrm{sep},t}\) measures the error in
the estimated inter-target separation. The implemented reward is
\begin{align}
r_t
=
-
\left(
e_{\mathrm{tar},t}^{\,2}
+
\lambda_{\mathrm{sep}}
e_{\mathrm{sep},t}^{\,2}
\right).
\label{eq:conference_reward}
\end{align}

The dimensionless coefficient \(\lambda_{\mathrm{sep}}\) controls the
relative importance of the separation error. For example,
\(\lambda_{\mathrm{sep}}=0.5\) gives the squared separation error half the
weight of the squared target-angle RMSE, whereas
\(\lambda_{\mathrm{sep}}=1.5\) gives it \(50\%\) more weight. The latter is
used to emphasize correct separation of the two targets during training. This reward is available only in the training simulator
and is not required during deployment.
Because the task terminates after one dwell, SAC maximizes

\begin{align}
J(\pi)
=
\mathbb{E}
\left[
r_t
+
\alpha
\mathcal{H}
\left(
\pi(\cdot|\mathbf{s}_t)
\right)
\right],
\label{eq:conference_sac_objective}
\end{align}

where \(\pi\) is the stochastic policy,
\(\mathcal{H}(\pi(\cdot|\mathbf{s}_t))\) is its entropy, and
\(\alpha\) is the entropy-temperature coefficient. Standard SAC
training uses one actor, two critic networks, target critics, and
an experience replay buffer.

After offline convergence, the actor is frozen. During deployment,
the radar forms the state in \eqref{eq:conference_state}, the actor
generates the five-dimensional structured action, and this action
is expanded into the RIS phase mask applied during the next radar
dwell. No true target angles, channel labels, or online gradient
updates are required.

\section{Numerical Results}
\label{sec:results}

The simulations consider two closely spaced vital-sign targets observed by a
$28~\mathrm{GHz}$ monostatic monopulse radar. Both the receive array and the
RIS comprise $40\times40$ half-wavelength-spaced elements. Unless otherwise
stated, the targets are located at a range of $15~\mathrm{m}$ and an elevation
of $20^{\circ}$, with an azimuth separation of $0.3^{\circ}$. Each radar dwell
contains $30$ pulses transmitted at a PRF of $40~\mathrm{Hz}$, corresponding
to a duration of $0.75~\mathrm{s}$. The nominal post-dwell sum-beam SNR is
$5~\mathrm{dB}$, the direct path is attenuated by $10~\mathrm{dB}$, and the
ghost response has a $-9~\mathrm{dB}$ amplitude ratio and a $4^{\circ}$
angular offset.

The monopulse sum and azimuth-difference channels coherently process all
$1600$ receive elements. Since elevation is fixed, the planar-array response
is evaluated using an equivalent separable implementation. The SAC actor and
two critic networks each contain two hidden layers of $96$ neurons. The policy
is initialized using $2500$ warm-start scenes over $35$ epochs and refined
through $3500$ SAC updates. Each operating point is averaged over $100$
matched Monte Carlo trials, with identical scenes, calibration errors, and
noise samples used across all compared methods.

\begin{table}[t]
\centering
\caption{Principal Simulation Parameters}
\label{tab:conference_parameters}
\resizebox{\columnwidth}{!}{%
\begin{tabular}{ll}
\toprule
\textbf{Parameter} & \textbf{Value} \\
\midrule
Carrier frequency & $28~\mathrm{GHz}$ \\
Receive array / RIS & $40\times40$ / $40\times40$ elements \\
Element spacing & $\lambda/2$ \\
Target range / elevation & $15~\mathrm{m}$ / $20^{\circ}$ \\
Nominal azimuth separation & $0.3^{\circ}$ \\
Pulses per dwell / PRF & $30$ / $40~\mathrm{Hz}$ \\
Dwell duration & $0.75~\mathrm{s}$ \\
FMCW bandwidth / chirp duration & $2~\mathrm{GHz}$ / $100~\mu\mathrm{s}$ \\
FMCW samples / range resolution & $512$ / $7.5~\mathrm{cm}$ \\
Nominal SNR & $5~\mathrm{dB}$, post-dwell sum beam \\
Direct-path attenuation & $10~\mathrm{dB}$ \\
Ghost ratio / angular offset & $-9~\mathrm{dB}$ / $4^{\circ}$ \\
Actor and critic hidden layers & $2\times96$ \\
Warm-start scenes / epochs & $2500$ / $35$ \\
SAC updates & $3500$ \\
Learning rate / batch size & $3\times10^{-4}$ / $128$ \\
Monte Carlo trials per point & $100$ \\
\bottomrule
\end{tabular}%
}
\end{table}

The proposed SAC controller is compared with no-RIS monopulse processing, an
FMCW range-angle FFT baseline, random-phase RIS, direction-only RIS,
target-calibrated RIS, full calibrated RIS, and a nondeployable exact-channel
reference. For the FMCW baseline, an unresolved target pair is assigned a zero
estimated separation. Performance is evaluated using the mean absolute error
of the estimated inter-target azimuth separation, and paired two-sided
$t$-tests compare SAC with the full calibrated controller.

\subsection{Angular-Separation Accuracy and SNR Robustness}

\begin{figure}[t]
    \centering
    \includegraphics[width=.85\columnwidth]
    {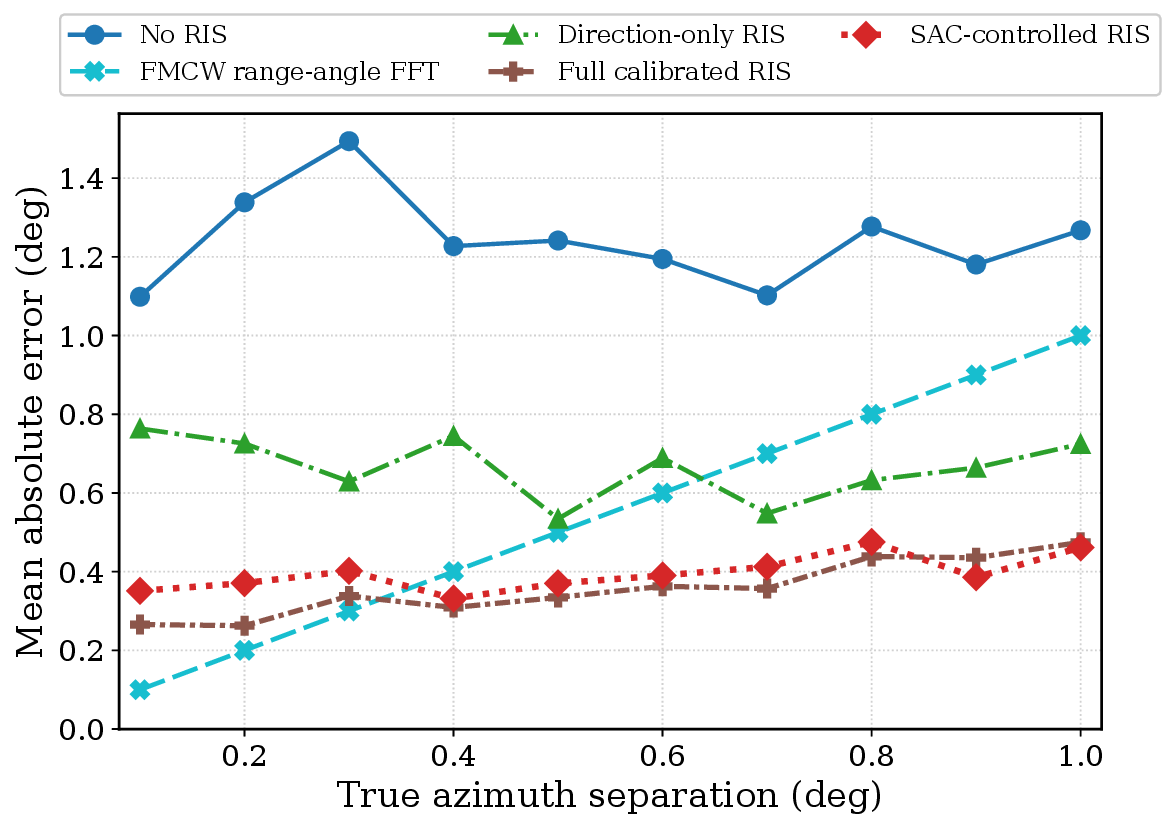}
    \caption{Mean estimated azimuth separation versus the true target
    separation at $5~\mathrm{dB}$ SNR, averaged over $100$ matched Monte Carlo
    trials. The diagonal line denotes ideal estimation. The FMCW curve
    remains at zero estimated separation because the target pair is
    unresolved throughout the tested interval.}
    \label{fig:fig2_error_vs_separation}
\end{figure}

As shown in Fig. \ref{fig:fig2_error_vs_separation}, at the nominal $0.3^{\circ}$ separation and $5~\mathrm{dB}$ SNR,
conventional monopulse processing without an RIS produces a separation MAE of
$1.494^{\circ}$, while the random-phase RIS gives
$1.491^{\circ}$. The FMCW range-angle FFT baseline yields an MAE
of $0.300^{\circ}$ and a $0\%$ resolved-pair rate because it detects only one
angular peak in all $100$ trials. Direction-only and target-calibrated RIS
control reduce the error to $0.629^{\circ}$ and
$0.522^{\circ}$, respectively. The full calibrated controller achieves
$0.338^{\circ}$, whereas the proposed SAC controller reaches
$0.402^{\circ}$. The exact-channel reference gives
$0.334^{\circ}$.

The SAC controller reduces the nominal MAE by $73.1\%$ relative to
no-RIS monopulse processing and by $36.1\%$ relative to direction-only
RIS control. However, the full calibrated controller is $18.9\%$ better
than SAC at this operating point, with a paired two-sided result of
$p=4.41\times10^{-2}$. Although the FMCW failure penalty of
$0.300^{\circ}$ is numerically below the SAC MAE, its resolved-pair rate is
$0\%$; therefore, it does not provide two-target angular localization. 

\begin{figure}[t]
    \centering
    \includegraphics[width=\columnwidth]
    {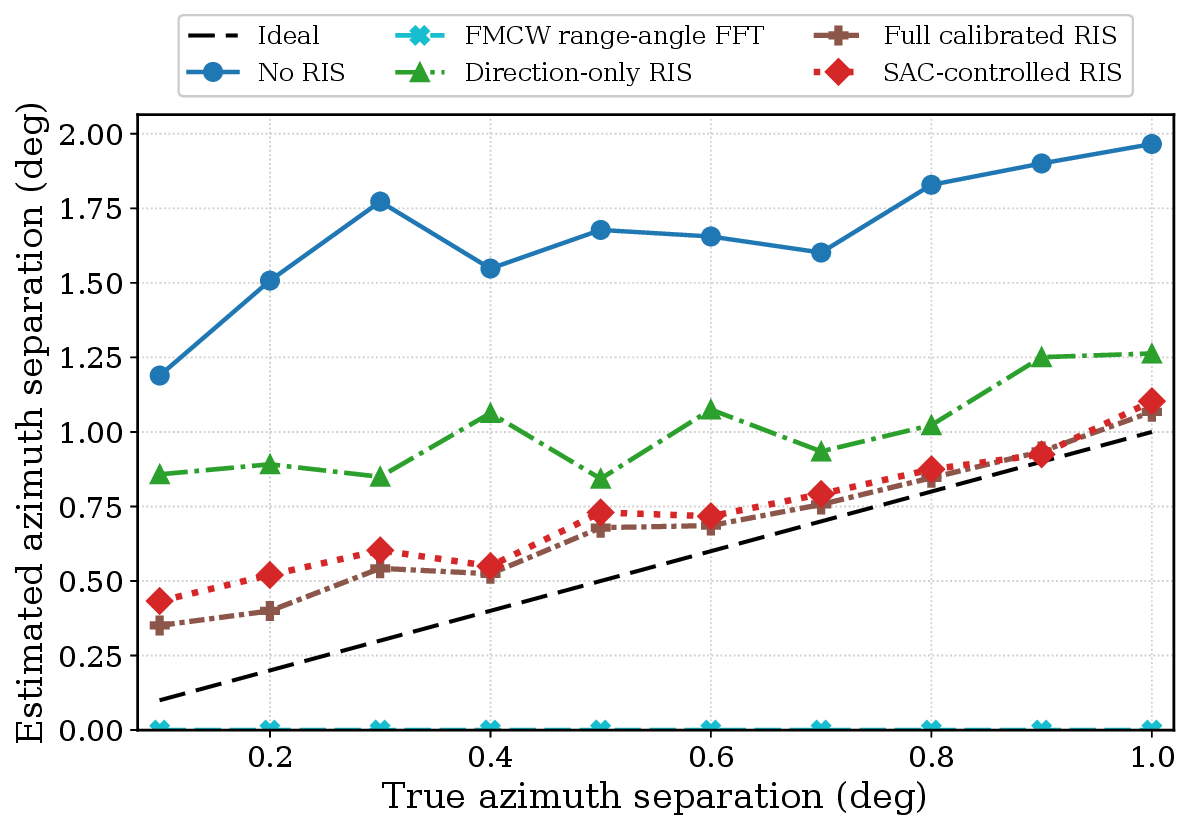}
    \caption{Mean estimated azimuth separation versus the true target
    separation at $5~\mathrm{dB}$ SNR, averaged over $100$ matched Monte Carlo
    trials. The diagonal line denotes ideal estimation. The FMCW curve
    remains at zero estimated separation because the target pair is
    unresolved throughout the tested interval.}
    \label{fig:estimated_vs_true_separation}
\end{figure}

Figure~\ref{fig:estimated_vs_true_separation} compares the mean estimated
azimuth separation with the true separation from $0.1^{\circ}$ to
$1^{\circ}$. The diagonal ideal line represents exact separation
estimation. The conventional FMCW estimate remains at $0^{\circ}$ over
the entire sweep because the processor does not resolve the target pair. The
no-RIS method strongly overestimates the separation, with its mean estimate
increasing from $1.189^{\circ}$ at a true separation of $0.1^{\circ}$
to $1.966^{\circ}$ at $1^{\circ}$. Direction-only RIS control reduces
this bias but estimates $0.858^{\circ}$ and $1.263^{\circ}$ at those
two endpoints. Full calibration and SAC follow the ideal line more closely:
the full calibrated estimates increase from $0.351^{\circ}$ to
$1.067^{\circ}$, while the SAC estimates increase from
$0.433^{\circ}$ to $1.103^{\circ}$. At the nominal $0.3^{\circ}$
separation, their mean estimates are $0.542^{\circ}$ and
$0.603^{\circ}$, respectively. Thus, both optimized controllers
overestimate very small separations but approach the ideal relation as
the separation increases. The paired MAE analysis, which is not directly
plotted in this figure, shows that full calibration is significantly better at
$0.1^{\circ}$, $0.2^{\circ}$, and $0.3^{\circ}$; the methods are
statistically comparable at the remaining separation points.

\begin{figure}[t]
    \centering
    \includegraphics[width=\columnwidth]{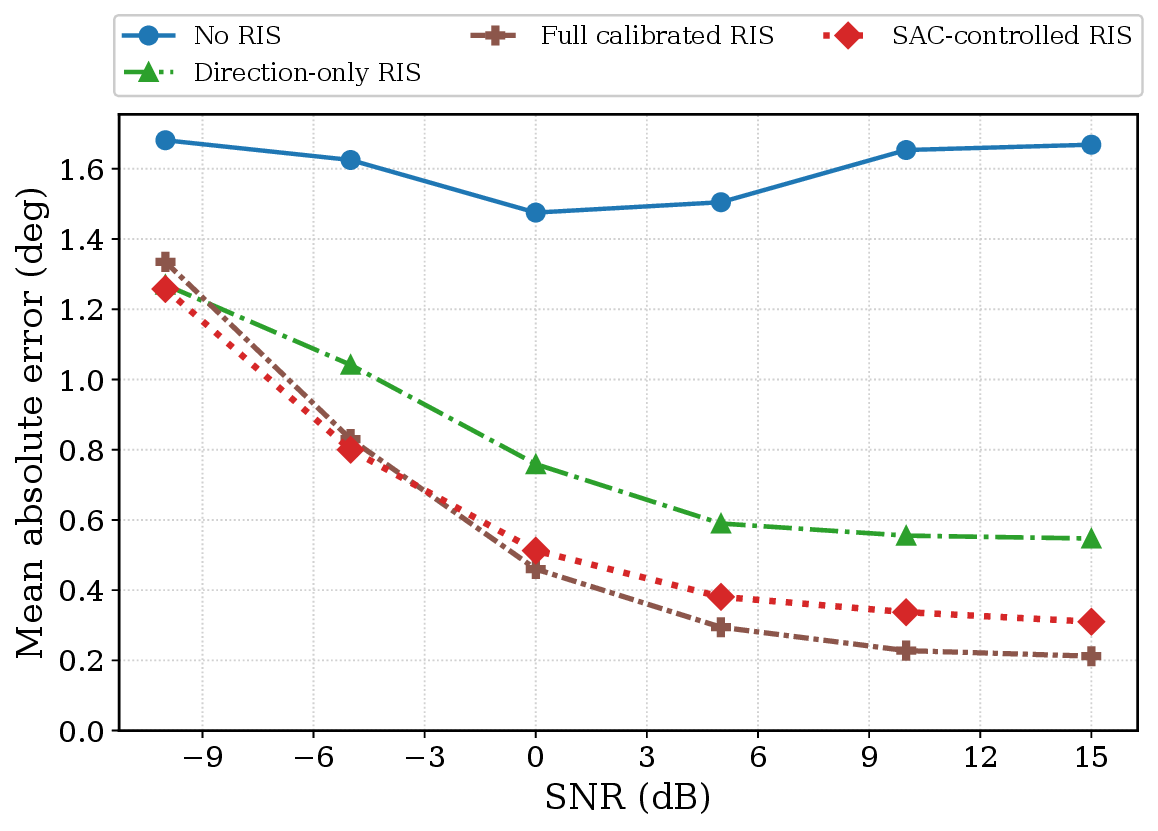}
    \caption{Mean absolute azimuth-separation error versus SNR for a fixed
    $0.3^{\circ}$ target separation, averaged over $100$ matched Monte Carlo
    trials.}
    \label{fig:error_vs_snr}
\end{figure}

Figure~\ref{fig:error_vs_snr} shows the SNR sweep from
$-10$ to $15~\mathrm{dB}$ at a fixed $0.3^{\circ}$ separation. With the
corrected noise normalization, the full calibrated error decreases from
$1.335^{\circ}$ at $-10~\mathrm{dB}$ to
$0.212^{\circ}$ at $15~\mathrm{dB}$, while the SAC error decreases from
$1.258^{\circ}$ to $0.310^{\circ}$. Direction-only RIS control
decreases from $1.270^{\circ}$ to $0.547^{\circ}$. In contrast,
the no-RIS result remains between $1.475^{\circ}$ and $1.681^{\circ}$
because its coherent multipath bias dominates after the thermal-noise
contribution is reduced. SAC and full calibration are statistically comparable
at $-10$, $-5$, and $0~\mathrm{dB}$; full calibration is significantly better
at $5$, $10$, and $15~\mathrm{dB}$. The curves therefore show a
physically meaningful transition from a noise-limited region toward a
multipath and calibration error floor.

\subsection{Ghost-Strength Robustness}

\begin{figure}[t]
    \centering
    \includegraphics[width=\columnwidth]
    {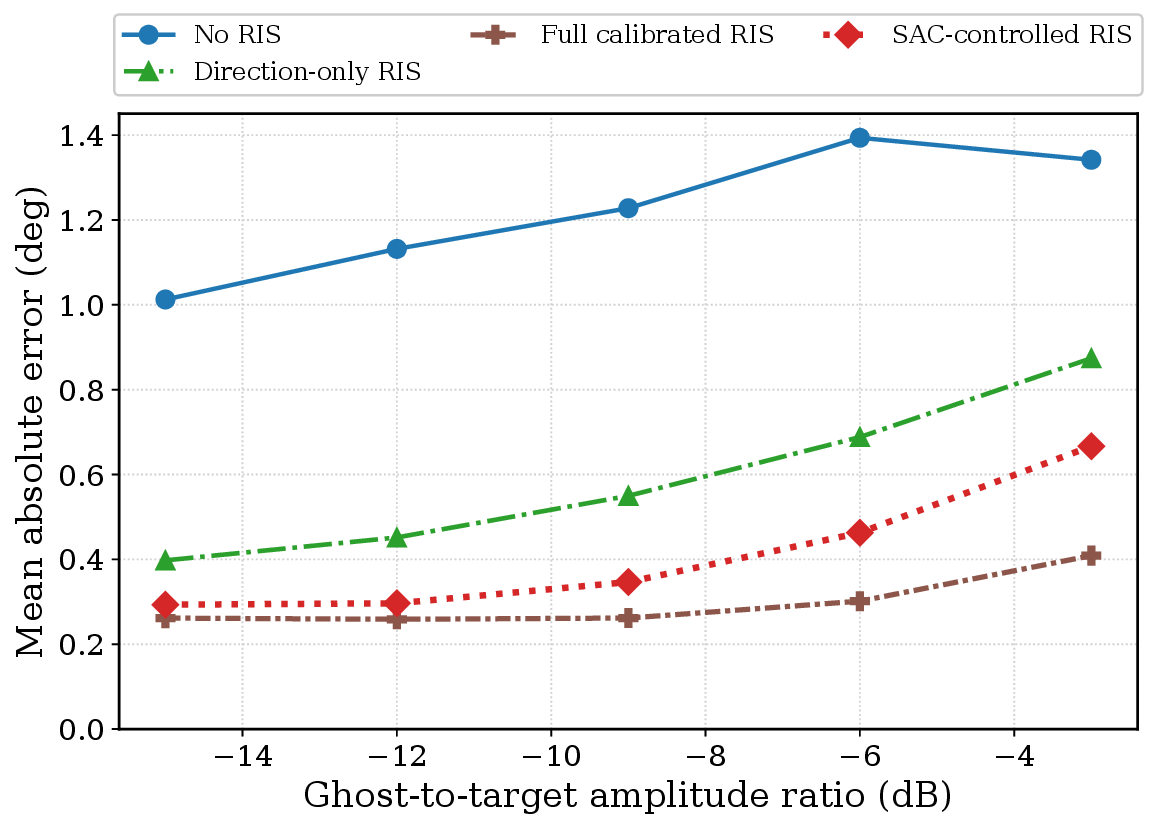}
    \caption{Mean absolute azimuth-separation error versus the ghost-to-target
    amplitude ratio over $100$ matched Monte Carlo trials.}
    \label{fig:error_vs_ghost_strength}
\end{figure}

Figure~\ref{fig:error_vs_ghost_strength} examines the ghost-to-target amplitude
ratio. As the ghost level increases from $-15$ to $-3~\mathrm{dB}$, the no-RIS
error changes from $1.013^{\circ}$ to $1.342^{\circ}$, and the
direction-only RIS error increases from $0.398^{\circ}$ to
$0.874^{\circ}$. The full calibrated controller increases from
$0.261^{\circ}$ to $0.409^{\circ}$, whereas SAC increases from
$0.293^{\circ}$ to $0.667^{\circ}$. The two methods are statistically
comparable at $-15$ and $-12~\mathrm{dB}$; full calibration is
significantly better at $-9$, $-6$, and $-3~\mathrm{dB}$. These results
show that the current SAC policy does not provide uniform superiority as
the coherent ghost becomes stronger.

\section{Conclusion}
\label{sec:conclusion}

This paper presented an SAC-controlled RIS-assisted monopulse radar for the
angular localization of two closely spaced vital-sign targets in multipath
environments. The framework employs a $40\times40$ receive array and a
$40\times40$ RIS, while the SAC policy is trained offline and applied during
deployment without access to the true target angles.
At the nominal $0.3^{\circ}$ separation and $5~\mathrm{dB}$ SNR, SAC reduces
the separation MAE from $1.494^{\circ}$ without an RIS and $0.629^{\circ}$
with direction-only RIS steering to $0.402^{\circ}$. The considered FMCW
baseline does not resolve the target pair over the tested separation interval.
Although SAC substantially improves over uncalibrated and direction-only RIS
methods, full calibration performs better at moderate-to-high SNR and under
strong ghost reflections. Future work will consider broader multipath
conditions, online policy refinement, and experimental validation.

\bibliographystyle{IEEEtran}
\bibliography{Monopulse_RIS}

\end{document}